\documentclass[preprint,aps,prc,nofootinbib]{revtex4}
\usepackage{epsfig}
\usepackage{graphicx}
\usepackage{amsfonts}
\usepackage{amsmath}
\begin{document}

\title{Erratum: Numerical values of the $f^F, f^D$ and $f^S$ coupling constants in SU(3)
invariant Lagrangian of the interaction of the vector-meson nonets with $1/2^+$ octet baryons
[Phys. Rev. C93, 055208 (2016)]}
\date{\today}

\medskip

\author{Cyril Adamu\v s\v cin}
\affiliation{Institute of Physics, Slovak Academy of Sciences,
Bratislava, Slovak Republic}

\author{Erik Barto\v s}
\affiliation{Institute of Physics, Slovak Academy of Sciences,
Bratislava, Slovak Republic}

\author{Stanislav Dubni\v cka}
\affiliation{Institute of Physics, Slovak Academy of Sciences,
Bratislava, Slovak Republic}

\author{Anna Z. Dubni\v ckov\'a}
\affiliation{Department of Theoretical Physics, Comenius University,
Bratislava, Slovak Republic}

\maketitle

   Some aspects of the presented determination of numerical values of the $f^F, f^D$ and $f^S$ coupling constants in the SU(3)
invariant interaction Lagrangian

\begin{eqnarray}
L_{VB\bar{B}}=\frac{i}{\sqrt{2}}f^F[\bar{B}^\alpha_\beta \gamma_\mu B^\beta_\gamma-\bar{B}^\beta_\gamma\gamma_\mu B^\alpha_\beta](V^\mu)^\gamma_\alpha+\nonumber\\
+\frac{i}{\sqrt{2}}f^D[\bar{B}^\alpha_\beta \gamma_\mu B^\beta_\gamma+\bar{B}^\beta_\gamma\gamma_\mu B^\alpha_\beta](V^\mu)^\gamma_\alpha+\nonumber\\
+\frac{i}{\sqrt{2}}f^S \bar{B}^\alpha_\beta \gamma_\mu B^\beta_\alpha\omega_0^\mu\nonumber
\end{eqnarray}
of the vector-meson nonet with $1/2^+$ octet baryons in \cite{abdd} were not correct.

   The crucial ingredient in that determination is an application of the $\omega-\phi$ mixing, which has been entered
into the procedure in \cite{abdd} two times. First in a derivation of expressions (51), (52), (55) and (56) for coupling
constants of vector-mesons with nucleons from the above-mentioned Lagrangian, and so, also in a derivation of the reverse
expressions (66), (67), (68) and (69), and then in a determination of the signs of the universal vector-meson coupling constants $f_\rho, f_\omega$ and $f_\phi$.

   Generally there are in literature the following four different physically acceptable forms of the $\omega-\phi$ mixing
\begin{eqnarray}
    1.\quad \phi&=&\omega_8 \cos \theta - \omega_0 \sin \theta\nonumber\\
            \omega&=&\omega_8 \sin \theta + \omega_0 \cos \theta \label{eq5}\\
    2.\quad \phi&=&-\omega_8 \cos \theta + \omega_0 \sin \theta\nonumber\\
            \omega&=&\omega_8 \sin \theta + \omega_0 \cos \theta \label{eq6}\\
    3.\quad \phi&=&-\omega_8 \cos \theta + \omega_0 \sin \theta\nonumber\\
            \omega&=&-\omega_8 \sin \theta - \omega_0 \cos \theta \label{eq7}\\
    4.\quad \phi&=&\omega_8 \cos \theta - \omega_0 \sin \theta\nonumber\\
            \omega&=&-\omega_8 \sin \theta - \omega_0 \cos \theta\,\label{eq8}
\end{eqnarray}
which manifest themselves in the four different forms of expressions for $f^F, f^D, f^S$

\begin{eqnarray}
    1.\quad f^F&=&\frac{1}{2}[f_{\rho NN}+\sqrt{3}(f_{\Phi NN} \cos \theta + f_{\omega NN} \sin \theta)]\nonumber\\
            f^D&=&\frac{1}{2}[3f_{\rho NN} -\sqrt{3}(f_{\Phi NN}\cos \theta + f_{\omega NN} \sin \theta)]\label{SU3cc1}\\
            f^S&=&\sqrt{2}(f_{\omega NN}\cos \theta - f_{\Phi NN}\sin \theta)\nonumber\\
    2.\quad f^F&=&\frac{1}{2}[f_{\rho NN}+\sqrt{3}(-f_{\Phi NN} \cos \theta + f_{\omega NN} \sin \theta)]\nonumber\\
            f^D&=&\frac{1}{2}[3f_{\rho NN} -\sqrt{3}(-f_{\Phi NN}\cos \theta + f_{\omega NN} \sin \theta)]\label{SU3cc2} \\
            f^S&=&\sqrt{2}(f_{\omega NN}\cos \theta + f_{\Phi NN}\sin \theta)\nonumber\\
    3.\quad f^F&=&\frac{1}{2}[f_{\rho NN}-\sqrt{3}(f_{\Phi NN} \cos \theta + f_{\omega NN} \sin \theta)]\nonumber\\
            f^D&=&\frac{1}{2}[3f_{\rho NN} +\sqrt{3}(f_{\Phi NN}\cos \theta + f_{\omega NN} \sin \theta)]\label{SU3cc3}\\
            f^S&=&-\sqrt{2}(f_{\omega NN}\cos \theta - f_{\Phi NN}\sin \theta)\nonumber\\
    4.\quad f^F&=&\frac{1}{2}[f_{\rho NN}+\sqrt{3}(f_{\Phi NN} \cos \theta - f_{\omega NN} \sin \theta)]\nonumber\\
            f^D&=&\frac{1}{2}[3f_{\rho NN} -\sqrt{3}(f_{\Phi NN}\cos \theta - f_{\omega NN} \sin \theta)]\label{SU3cc4}\\
            f^S&=&-\sqrt{2}(f_{\omega NN}\cos \theta + f_{\Phi NN}\sin \theta).\nonumber
\end{eqnarray}

   On the other hand an application of the same $\omega-\phi$ mixing configurations (\ref{eq5}), (\ref{eq6}),
(\ref{eq7}), (\ref{eq8}) leads to the rates of the overthrown
values of the universal vector-meson coupling constants with different signs as follows
\begin{eqnarray}
  1.\quad \frac{1}{f_\rho}:\frac{1}{f_\omega}:\frac{1}{f_\phi}&=&\sqrt{3}:\sin \theta:\cos \theta,\label{eq9}
\end{eqnarray}
which can be found e.g. in ref. \cite{perren} on p. 52 relation (A)
\begin{eqnarray}
  2.\quad \frac{1}{f_\rho}:\frac{1}{f_\omega}:\frac{1}{f_\phi}&=&\sqrt{3}:\sin \theta:-\cos \theta\label{eq10}\\\nonumber
\end{eqnarray}
to be found in ref. \cite{clcott} on p. 539
\begin{eqnarray}
  3.\quad \frac{1}{f_\rho}:\frac{1}{f_\omega}:\frac{1}{f_\phi}&=&\sqrt{3}:-\sin \theta:-\cos \theta \label{eq11}\\
  4.\quad \frac{1}{f_\rho}:\frac{1}{f_\omega}:\frac{1}{f_\phi}&=&\sqrt{3}:-\sin \theta:\cos \theta,\label{eq12}
\end{eqnarray}
and this last case can be found e.g. in ref. \cite{gas} on p. 446.

   All these relations can be explained by the following considerations.

   Starting e.g. from the $\omega-\phi$ mixing configuration (\ref{eq5})
and substituting explicitly
\begin{eqnarray}
   \omega_8 &=& \frac{1}{\sqrt{6}}(\bar u u +\bar d d -2 \bar s s)\label{eq13}\\\nonumber
   \omega_0 &=& \frac{1}{\sqrt{3}}(\bar u u +\bar d d + \bar s s)
\end{eqnarray}
and for the ideal mixing angle $\theta=35.3^0$
\begin{equation}
   \sin \theta = \sqrt \frac{1}{3}; \quad \cos \theta = \sqrt \frac{2}{3}, \label{eq14}
\end{equation}
one obtains
\begin{equation}
   \phi = -\bar s s\label{eq15}
\end{equation}
and
\begin{equation}
   \omega = +\frac{1}{\sqrt{2}}(\bar u u +\bar d d)\label{eq16}.
\end{equation}

   On the other hand the hadronic electromagnetic (EM) current
\begin{equation}
  J^h_\mu = \frac{2}{3}\bar u \gamma_\mu u -\frac{1}{3}\bar d \gamma_\mu d -\frac{1}{3}\bar s \gamma_\mu s,
\end{equation}
can be formally arranged to the shape
\begin{eqnarray*}
  J^h_\mu = \frac{1}{2}(\bar u\gamma_\mu u -\bar d\gamma_\mu d) + \frac{1}{6}(\bar u\gamma_\mu u +\bar d\gamma_\mu d)-\frac{1}{3}(\bar s\gamma_\mu s),
\end{eqnarray*}
and because

     $\frac{1}{\sqrt{2}}(\bar u \gamma_\mu u - \bar d \gamma_\mu d)=J^{\rho^0}_\mu$

     $ +\frac{1}{\sqrt{2}}(\bar u \gamma_\mu u + \bar d \gamma_\mu d)=J^\omega_\mu $ due to the sign "+" in (\ref{eq16})

     $-\bar s \gamma_\mu s=J^\phi_\mu $  due to the sign "-" in (\ref{eq15}),\\
are the $\rho^0-, \omega-, \phi -$ meson EM currents, respectively, the hadronic EM current acquires the following form

\begin{equation}
  J^h_\mu = \frac{1}{\sqrt{2}} J^{\rho^0}_\mu + \frac{1}{3\sqrt{2}} J^\omega_\mu +\frac{1}{3} J^\phi_\mu.\label{hemc1}
\end{equation}

   Now, if the results of the Kroll-Lee-Zumino paper \cite{klz}, that a linear combination of the neutral vector-meson  fields $\rho^0_\mu, \omega_\mu, \phi_\mu$
\begin{equation}
   J^h_\mu = \frac{m^2_{\rho^0}}{f_\rho}\rho^0_\mu + \frac{m^2_\omega}{f_\omega} \omega_\mu + \frac{m^2_\phi}{f_\phi} \phi_\mu\label{eq19},
\end{equation}
with the universal vector-meson coupling constants $f_\rho, f_\omega, f_\phi$, is proportional by some real constant $A$ to the
hadronic EM current (\ref{hemc1}), are taken into account, considering a dimension of the Dirac quark fields $u, d, s$ in (\ref{hemc1}), in the framework of
the natural units $\hbar=c=1$, to be m$^\frac{3}{2}$ and a dimension of the vector-meson fields in (\ref{eq19}) to be m$^1$,
the relations

\begin{equation}\label{eq20}
  \frac{1}{f_\rho} = A \frac{1}{\sqrt{2}}; \quad \frac{1}{f_\omega} = +A \frac{1}{3\sqrt{2}}; \quad \frac{1}{f_\phi} = +A\frac{1}{3},
\end{equation}
are found.

   Then from these last relations the rates
\begin{eqnarray*}
  \frac{1}{f_\rho}:\frac{1}{f_\omega}:\frac{1}{f_\phi}&=&\frac{1}{\sqrt{2}}:+\frac{1}{3\sqrt{2}}:+\frac{1}{3}\\
                                                                                 &=&\frac{\sqrt{6}}{\sqrt{2}}:+\frac{\sqrt{6}}{3\sqrt{2}}:+\frac{\sqrt{6}}{3}\\
                                                                                 &=&\sqrt{3}:+\frac{1}{\sqrt{3}}:+\sqrt{\frac{2}{3}}\\
                                                                                 &=&\sqrt{3}:+\sin\theta:+\cos\theta,\\
\end{eqnarray*}
are obtained, giving the signs of the universal vector-meson coupling constants $+f_\rho, +f_\omega, +f_\phi$.

   As a matter of fact the signs of the universal vector-meson coupling constants are specified already from the relations (\ref{eq20}),
however, a community of physicists prefers the relations (\ref{eq9})-(\ref{eq12}), in which the universal vector-meson coupling constants
are related to $\sin\theta$ and $\cos\theta$ where the angle $\theta$ is determined from the quadratic Gell-Mann-Okubo vector-meson mass formula, which provide
more realistic values of the latter to be in fair agreement with experimental evaluations. Therefore  we also favor a presentation of
the universal vector-meson coupling constants signs in the form (\ref{eq9})-(\ref{eq12}).

   In a like manner, starting from the $\omega-\phi$ mixing configuration (\ref{eq6}), and relations (\ref{eq13}) and (\ref{eq14}),
one obtains
\begin{equation}
   \phi = +\bar s s\label{eq21}
\end{equation}
and
\begin{equation}
   \omega = +\frac{1}{\sqrt{2}}(\bar u u +\bar d d)\label{eq22}.
\end{equation}
Then comparing the hadronic EM current to be multiplied by the real constant $A$
\begin{equation}
  J^h_\mu = \frac{1}{\sqrt{2}} J^{\rho^0}_\mu + \frac{1}{3\sqrt{2}} J^\omega_\mu -\frac{1}{3} J^\phi_\mu,\label{hemc2}
\end{equation}
where

    $J^{\rho^0}_\mu = \frac{1}{\sqrt{2}}(\bar u \gamma_\mu u - \bar d \gamma_\mu d)$

  $J^\omega_\mu = +\frac{1}{\sqrt{2}}(\bar u \gamma_\mu u + \bar d \gamma_\mu d)$ due to the sign "+" in (\ref{eq22})\\
  $J^\phi_\mu = +\bar s \gamma_\mu s$ due to the sign "+" in (\ref{eq21}),\\
with (\ref{eq19}), one obtains relations

\begin{equation}
  \frac{1}{f_\rho} = A\frac{1}{\sqrt{2}}; \quad \frac{1}{f_\omega} = +A\frac{1}{3\sqrt{2}}; \quad \frac{1}{f_\phi} = -A\frac{1}{3}\label{24}
\end{equation}
and from them the rates (\ref{eq10}), giving the following signs of universal vector-meson coupling constants $+f_\rho, +f_\omega, -f_\phi$.

   Again, starting from the $\omega-\phi$ mixing configuration (\ref{eq7}), and relations (\ref{eq13}) and (\ref{eq14}),
one obtains
\begin{equation}
   \phi = +\bar s s\label{eq25}
\end{equation}
and
\begin{equation}
   \omega = -\frac{1}{\sqrt{2}}(\bar u u +\bar d d)\label{eq26}.
\end{equation}
Then comparing the hadronic EM current to be multiplied by the real constant $A$
\begin{equation}
  J^h_\mu = \frac{1}{\sqrt{2}} J^{\rho^0}_\mu - \frac{1}{3\sqrt{2}} J^\omega_\mu -\frac{1}{3} J^\phi_\mu,\label{hemc3}
\end{equation}
where

    $J^{\rho^0}_\mu = \frac{1}{\sqrt{2}}(\bar u \gamma_\mu u - \bar d \gamma_\mu d)$

  $J^\omega_\mu = -\frac{1}{\sqrt{2}}(\bar u \gamma_\mu u + \bar d \gamma_\mu d)$ due to the sign "-" in (\ref{eq26})\\
  $J^\phi_\mu = +\bar s \gamma_\mu s$ due to the sign in (\ref{eq25}),\\
with (\ref{eq19}), one obtains relations

\begin{equation}
  \frac{1}{f_\rho} = A\frac{1}{\sqrt{2}}; \quad \frac{1}{f_\omega} = -A\frac{1}{3\sqrt{2}}; \quad \frac{1}{f_\phi} = -A\frac{1}{3}
\end{equation}
and from them the rates (\ref{eq11}), giving the following signs of universal vector-meson coupling constants $+f_\rho, -f_\omega, -f_\phi$.

   Finally, starting from the $\omega-\phi$ mixing configuration (\ref{eq8}), and relations (\ref{eq13}) and (\ref{eq14}),
one obtains
\begin{equation}
   \phi = -\bar s s\label{eq29}
\end{equation}
and
\begin{equation}
   \omega = -\frac{1}{\sqrt{2}}(\bar u u +\bar d d)\label{eq30}.
\end{equation}
Then comparing the hadronic EM current to be multiplied by the real constant $A$
\begin{equation}
  J^h_\mu = \frac{1}{\sqrt{2}} J^{\rho^0}_\mu - \frac{1}{3\sqrt{2}} J^\omega_\mu +\frac{1}{3} J^\phi_\mu,\label{hemc4}
\end{equation}
where

    $J^{\rho^0}_\mu = \frac{1}{\sqrt{2}}(\bar u \gamma_\mu u - \bar d \gamma_\mu d)$

  $J^\omega_\mu = -\frac{1}{\sqrt{2}}(\bar u \gamma_\mu u + \bar d \gamma_\mu d)$ due to the sign "-" in (\ref{eq30})\\
  $J^\phi_\mu = -\bar s \gamma_\mu s$ due to the sign "-" in (\ref{eq29}),\\
with (\ref{eq19}), one obtains relations

\begin{equation}
  \frac{1}{f_\rho} = A\frac{1}{\sqrt{2}}; \quad \frac{1}{f_\omega} = -A\frac{1}{3\sqrt{2}}; \quad \frac{1}{f_\phi} = +A\frac{1}{3}
\end{equation}
and from them the rates (\ref{eq12}), giving the following signs of universal vector-meson coupling constants $+f_\rho, -f_\omega, +f_\phi$.

   The signs of the universal vector-meson coupling constants $f_\rho, f_\omega, f_\phi$ are very important to be known, as the numerical
values of these constants are regularly estimated from the experimental values \cite{olive} of the vector-meson lepton widths
by means of the formula

\begin{equation}
  \Gamma(V \to e^+e^-)=\frac{\alpha^2 m_V}{3}{\left(\frac{f^2_V}{4\pi}\right)}^{-1}\label{lepwidth},
\end{equation}
in which $f_V$ is contained in a quadratic form.

   From (\ref{SU3cc1}), (\ref{SU3cc2}), (\ref{SU3cc3}), (\ref{SU3cc4}) it seems at first sight that numerical values
of the coupling constants $f^F, f^D, f^S$ have to depend on the choice of the $\omega-\phi$ mixing version.

   However, if in (\ref{SU3cc1}), (\ref{SU3cc2}), (\ref{SU3cc3}), (\ref{SU3cc4}) the coupling constants of vector-mesons
with nucleons are determined e.g from $(f^{(1)}_{\rho N N}/f_\rho), (f^{(1)}_{\omega N N}/f_\omega), (f^{(1)}_{\phi N N}/f_\phi)$,
to be found in a fitting procedure of all existing data on nucleon EM structure, by means of the signs of
$f_\rho, f_\omega, f_\phi$ following from (\ref{eq9}), (\ref{eq10}), (\ref{eq11}), (\ref{eq12}), as it is demonstrated above, then from all four
different expressions in (\ref{SU3cc1}), (\ref{SU3cc2}), (\ref{SU3cc3}), (\ref{SU3cc4}) one obtains the same numerical
values for $f^F_1, f^D_1, f^S_1$ as follows
\begin{eqnarray}
   f^F_1=5.414;\quad f^D_1=-1.699;\quad f^S_1=42.916\,.\label{eq35}
\end{eqnarray}

   By means of a similar procedure one can find also numerical values of all other coupling constants under
consideration
\begin{eqnarray}
   f^F_2&=&7.626;\quad f^D_2=21.088;\quad f^S_2=-7.111;\nonumber\\
   f^{F'}_1&=&8.343;\quad f^{D'}_1=12.498;\quad f^{S'}_1=-7.858; \label{eq36}\\
   f^{F'}_2&=&-30.450;\quad f^{D'}_2=-5.271;\quad f^{S'}_2=-18.614\,.\nonumber
\end{eqnarray}

   In the paper \cite{abdd} for the numerical evaluation of the vector-meson-nucleon coupling
constants from the $(f_{\rho N N}/f_\rho), (f_{\omega N N}/f_\omega), (f_{\phi N N}/f_\phi)$ we have applied the signs of the
universal vector-meson coupling constants (\ref{eq10}) (to be inspired by Close and Cottingham in \cite{clcott}),
which moreover have been combined with expressions for $f^F, f^D, f^S$
\begin{eqnarray}
&&f^F=\frac{1}{2}\left[f_{\rho NN}+ \sqrt{3} \left(f_{\phi NN}\cos \theta-f_{\omega NN}\sin \theta\right)\right] \\
&&f^D=\frac{1}{2}\left[3f_{\rho NN} -\sqrt{3}\left(f_{\phi NN} \cos \theta-f_{\omega NN}\sin \theta\right) \right] \nonumber \\
&&f^S=\sqrt{2}\left(f_{\omega NN}\cos \theta+f_{\phi NN}\sin \theta\right) \nonumber
\end{eqnarray}
to be generated by the physically non-acceptable form of the $\omega-\phi$ mixing
\begin{eqnarray}
\quad \phi&=&\omega_8 \cos \theta + \omega_0 \sin \theta\nonumber\\
\omega&=-&\omega_8 \sin \theta + \omega_0 \cos \theta \\
\end{eqnarray}
used by Gasiorovicz \cite{gas} on p.325, leading to the numerical values of $f^F, f^D, f^S$ differing
from the correct values presented in this Erratum.


\begin{thebibliography}{99}

\bibitem{abdd} C.Adamuscin, E.Bartos, S.Dubnicka and A.Z.Dubnickova, Phys. Rev. C93, 055208 (2016)

\bibitem{perren} J. P.Perez-y-Jorba and F. M.Renard, Phys. Reports 31C, 1 (1977)

\bibitem{clcott} F.E.Close and W.N.Cottingham: $e^+e^-$ Annihilation in \textit{Electromagnetic Interactions of Hadrons, Vol.2
                 Editors: A.Donnachie, G.Shaw, Originally published by Plenum Press, New York in 1978}

\bibitem{gas} S.Gasiorowicz: Elementary particle physics, John Wiley \& Sons, Inc. New York, 1966

\bibitem{klz} N. M.Kroll, T. D.Lee and B. Zumino, Phys. Rev. 157, 1376 (1967)

\bibitem{olive} C. Patrignani et al.(Particle Data Group), Chin. Phys. C40, 100001 (2016)



\end{thebibliography}
\end{document}